# Virtual Reflexes

Catholijn M. Jonker,[1] Joost Broekens,[1] Aske Plaat[2]

[1] Interactive Intelligence
Delft University of Technology
Mekelweg 4
The Netherlands
`c.m.jonker@tudelft.nl,`
`joost.broekens@gmail.com`

[2] Tilburg School of Humanities
Tilburg University
Warandelaan 2
The Netherlands
`aske.plaat@gmail.com`

**Abstract.** Virtual Reality is used successfully to treat people for regular phobias. A new challenge is to develop Virtual Reality Exposure Training for social skills. Virtual actors in such systems have to show appropriate social behavior including emotions, gaze, and keeping distance. The behavior must be realistic and real-time. Current approaches consist of four steps: 1) trainee social signal detection, 2) cognitive-affective interpretation, 3) determination of the appropriate bodily responses, and 4) actuation. The "cognitive" detour of such approaches does not match the directness of human bodily reflexes and causes unrealistic responses and delay. Instead, we propose *virtual reflexes* as concurrent sensory-motor processes to control virtual actors. Here we present a virtual reflexes architecture, explain how emotion and cognitive modulation are embedded, detail its workings, and give an example description of an aggression training application.

## 1 Introduction

The idea of using virtual reality to treat regular phobia—such as fear of heights, or flying—relies on the success with which we can immerse the patient in a virtual situation. Patients should experience the same emotions and stress as in a corresponding real life situation. During the session, patients (trainees) learn to become aware of their own bodily and emotional responses. Based on that awareness they learn how to control their own body and emotions. For training *social* skills, the VR setting has to be such that the emotions and intentions that trainees would attribute to a real person are now attributed to a Virtual Character (VC). In order for trainees to be able to learn to control themselves and change their behavior, they first have to correctly attribute emotions and intentions to the other person (Meichenbaum, 1994).

Different computational modeling approaches can be followed to produce appropriate social virtual character behavior. The high-level approach is to deduce social concepts from low-level trainee observations; then assemble the inputs of the various sensors and represent them conceptually, and use these conceptual emotions in an interpretation and reasoning process to produce responses at the conceptual level; finally translate these conceptual responses to muscle actuations. In contrast to this high-level approach, in this paper we propose a low-level approach. Our low-level approach is based on the idea of virtual reflexes, in which observations directly cause VC muscle actuations *without intermediate cognitive processing*. Based on various primitive inputs, the muscle actuations are immediate and create bodily reactions. This generates fluent and fast responses in the VC. Perceived emotions emerge out of the interaction between sensory and motor information. Although no cognitive intermediate processing takes place, cognition and affect do modulate the sensory motor loops. As happens in the human body, various virtual reflexes can occur simultaneously. Therefore, we model the set of virtual reflexes as concurrent subsystems.

The idea that social signals emerge from virtual reflexes is motivated intuitively, practically and theoretically. First, when engaging in day-to-day activities, people do whatever it is they are doing, in unthinking response to the "moment-to-moment local forces acting upon them" (Wakefield & Dreyfus 1991, p. 263). This is closer to a reflex-based approach than to a reasoning-based one. Second, virtual reflexes provide the speed necessary for realistic social interaction (Magnenat-Thalmann & Thalmann, 2005). Third, reflex-based control corresponds to theories on embodied cognition and affect (Wilson, 2002; Ziemke, 2003). The contributions of this paper are as follows:

- We introduce an architecture for virtual reflexes.
- We link the architecture to neuropsychological theories on emotion & cognition.
- We formalize part of the reflexes in a virtual aggression training case study.

In section 2 we discuss related work on interaction with virtual characters. Section 3 presents the virtual reflex architecture. The neurological and psychological basis is presented in Section 4. Section 5 presents a virtual aggression-training case study.

## 2    Related work

Virtual reality techniques have been shown to have significant therapeutic value, and in particular, VR stress inoculation training works well in various settings (Serino et al., 2013). Virtual reality exposure therapy (Emmelkamp et al., 2001; Krijn et al., 2004; Powers & Emmelkamp, 2008) has been shown to be as effective as in vivo (real-world) exposure therapy (Powers & Emmelkamp, 2008). Popovic et al (2009) present a Stress Inoculation system using an interactive VR system. Virtual reality systems have yielded positive training results (Core et al., 2006; Hays et al., 2010; Parsons & Mitchell, 2002; Spek, 2011; Broekens, et al., 2012; Kim et al., 2009; Zeng et al., 2009), and can also be used for personality assessment (Tekofsky et al., 2013).

An important challenge in interactive story telling (Cavazza et al., 2002; Theune et al., 2003; Zwaan et al., 2012) is that autonomous behavior of virtual characters must be consistent with the storyline. This challenge is similar, but on a different time scale, to the challenge in VC control where automatic reflexive behaviors must be consistent with deliberative and reflective behavior.

The real-time aspect of emotion modeling has been addressed in, e.g., the work on autonomous real-time sensitive artificial listeners (D'Mello et al., 2007; Pantic & Rothkrantz, 2003; Schroder et al., 2012; Thiebaux et al., 2008), the work on back-channel communication (Cafaro et al., 2012; Heylen et al., 2005; Sevin et al., 2010), to create "rapport" between virtual agents and humans (Gratch et al., 2007; Huang et al., 2011; Finkelstein et al., 2012), and in computational models of coping (Marsella et al., 2009) and synthetic emotion generation for games (Popescu et al., 2013). The challenge of generating real-time behavior has motivated (Brooks, 1999) to develop the subsumption architecture. Our architecture is inspired by Brooks' work in the sense that it consists of multiple subsystems running concurrently.

## 3     Virtual Reflex Architecture

The basis of our architecture is "the need for speed" of immersive virtual emotions, and the uncoupling of various sensor-actuator channels. Behavior is generated by reflex nodes that dynamically couple sensory input and motor output. To cope with high-level influences on behavior, such as training scenarios, activity of these nodes is modulated by cognitive and emotional factors (see figure 1). Each reflex node can be seen as a small control node that influences body parts. Its output is based on whether its sensory input deviates from a preset baseline, much like drives would need to be met in a homeostatic approach (Cañamero, 2005). Upon deviation, three things happen concurrently. First, the deviation triggers activation of the body parts coupled to the reflex node. Second, the deviation has an effect on the emotional state. We envision a Pleasure–Arousal–Dominance (PAD) representation (Mehrabian, 1980) of the emotional state (Emotion, in figure 1). Third, the deviation is available for cognition to reason upon. The emotional state is simply a correlate of the aggregated deviations from the drives, and as such "setting" the emotional state will also bias the drives towards a different baseline. This provides a natural and behaviorally grounded mechanism to model the influence of emotion on behavior, and also the emergence of emotion out of behavior and reflexes (Cañamero, 2005). The cognitive model operates on sensory-motor primitives, as the information it gets is not the raw sensory information but the deviation and drive-based control effect following from the sensory information. In our architecture, cognition is grounded in sensory-motor representations, which is in line with embodied cognition approaches (Wilson, 2002). Cognition influences behavior by modulating the virtual reflex decision node activities and parameters, just like emotion does. Emotion and cognition thus follow from and operate on reflexes in similar ways. In our approach the emotional state is simply a different representation of the emergent pattern of reflex activities, while cognition can hold any processing mechanism, as long as it takes as input reflex-node activity and it outputs reflex-node biases. The link from the body of the virtual character to the proprioceptive part of the perception system is in line with the body loop of Damasio (1999). It allows the virtual character to perceive, and respond to, its own actions.

The virtual character's immediate responses are controlled by the virtual reflex-loop, while the agent's high-level decisions are modeled as cognitive biases to the virtual reflex decision nodes. Scenario-based cognitive decision-making influences the behavior of the virtual character. In essence, cognitive processing biases the reflexes. In this way the architecture also allows emotional coping (Folkman & Lazarus, 1990) by means of influencing the reflexive behaviors, which, in turn, influence the emotional state. This is a natural way of modeling coping, as this grounds the coping process in the actual "physiology" of the virtual character instead of simply influencing the representation of the emotional state.

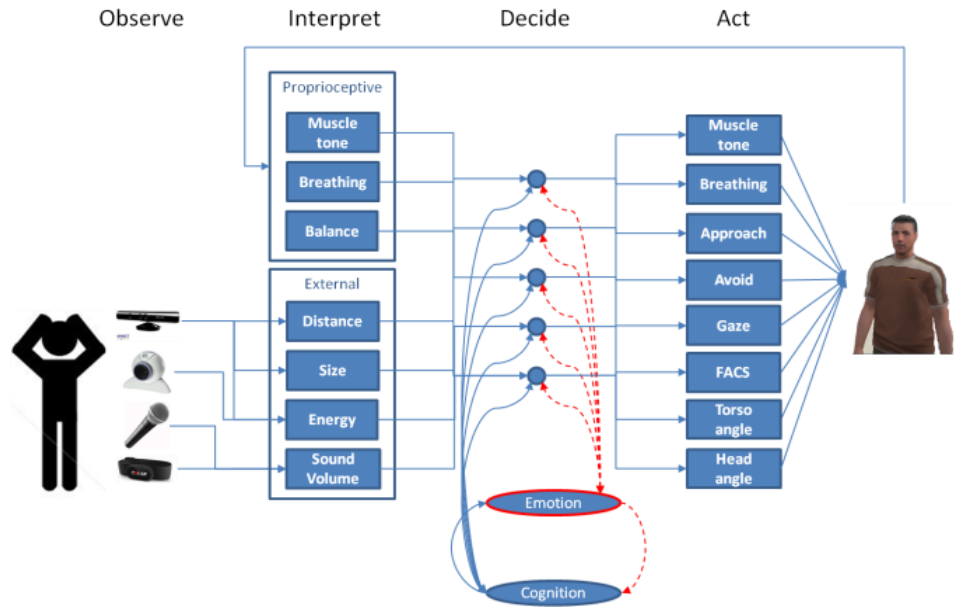

**Fig. 1.** Virtual Reflex architecture showing for each sensor-subsystem different reflex nodes that process the signals and yield appropriate responses. Note that Virtual Reflexes are modulated by cognition and emotion.

We assume that the VC has a body integrity model allowing it to, e.g., walk forward when the torso is pitched to the front, in order to keep balance. Candidate solutions are physics based models (Faloutsos et al., 2001) allowing individual influences on body parts with effects propagating through the complete body.

## 4   Neuropsychological Basis

In our approach behavior generation is the cognitive (and affective) modulation of automatic reflex processes (Berthoz, 2002). Our approach is compatible with embodied cognition theory (Wilson, 2002; Ziemke, 2003), where the basic premise is that thought is tightly coupled to behavior and the representations used for behavior.

Note that the idea of multiple levels of increasingly complex processing involved in affective responses is not new. For example, LeDoux (1995; 1996) describes a high and a low route to emotion processing. The low route is "quick and dirty" and evaluates stimuli in a fast but inaccurate way, while the high route is cortical and evaluates stimuli in a slow but detailed way. Scherer (2001) also considers appraisal as a process of multi-level sequential checking with lower levels triggering activity at higher levels. Each level typically involves more complex information processing, and is only activated to the full extent when that is needed based on simpler appraisals. For example, appraising stimulus relevance is done first based on suddenness, a simple stimulus-based appraisal, and when relevance is high this triggers implication-related appraisals such as goal-conduciveness, a more cognitive-based appraisal. This model of appraisal can be computationally integrated with other models that assume appraisal is layered from simple to complex processing (Broekens et al., 2008).

The ideas of thinking fast and slow (Kahneman, 2011) contain the same basic idea, i.e., that there are multiple concurrent processes for different aspects of behavior and reasoning. The fastest behaviors we have are reflex behaviors, developed early in our evolution, and essential for survival. Body language is an important part of social interaction (Argyle, 2009; Kendon, 1990), and is recognized as expressions of emotions. Mimicry is an essential part of the human social repertoire that is inexorably bound up to basic social processes of empathy, bonding, and in-group formation (Kavanagh, 2013). In contrast, higher order cognitive appraisal processes take more time, e.g., (Ortony, Clore & Collins, 1988), and explain how we appraise our progress with respect to our social and personal goals.

Our approach is also inspired by Damasio's somatic marker hypothesis (1999). Somatic markers are the affective counterparts of situational representations in terms of sensory-motor activity. Such markers get triggered by activities and thoughts, and can bias behavior and thought at the same time (for example, during decision making). Our emotional state emerges from the activity of the reflex nodes; it is grounded in sensory-motor activity. The state itself influences these nodes, as the relation between emotion and reflex is bidirectional. If the agent is in a high arousal state, this biases the reflexes towards high energy. This, in turn, biases cognition to trigger those associations that are related to high energetic reflex behavior, closing the loop from emotion to cognition through sensory-motor representations. In our approach, "feelings are mental experiences of body states" (Damasio & Carvalho, 2013).

## 5   Case: Aggression De-escalation

As an example case we explain how the Virtual Reflex architecture can form the basis for a Virtual Reality training system for coping with verbal aggressive situations, see figure 2. The system confronts a trainee in virtual reality with a verbal aggressor.

During the training a virtual character approaches the human trainee with verbal aggressive behavior and subsequently reacts in real-time to the actions and behaviors of the human trainee. A computational escalation model determines how this aggression is portrayed by the behavior of the virtual character. The actual virtual character's behaviors are generated in real-time based on (1) the output of this escalation model, (2) the behavioral responses of the trainee, and (3) a scenario setting given by aggression trainers. In this paper we focus on the VC behavior generation component by means of a virtual reflex architecture, not on the scenario or escalation model.

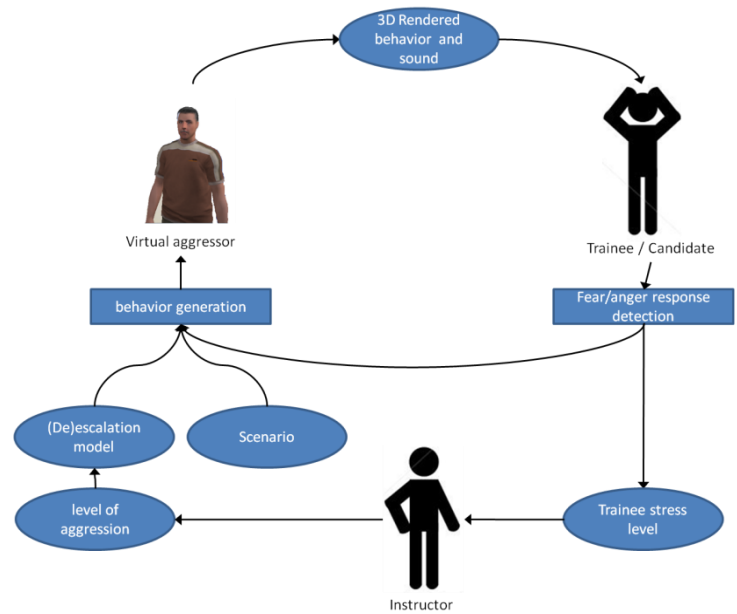

**Fig. 2.** Aggression De-escalation system overview.

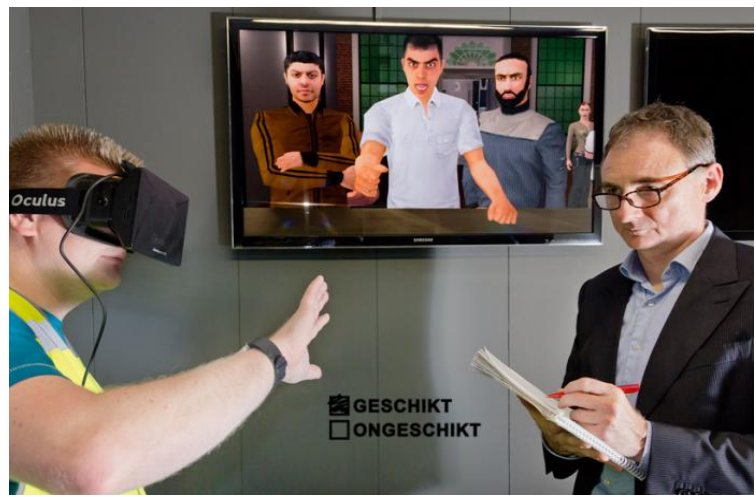

**Fig. 3.** An impression of the VR System

### 5.1 Behavior Generation and Virtual Reflex-loops

We now explain how the architecture presented in Section 2 is used for behavior generation in virtual aggression training. The complete system will contain a VC emotional state, dialogue, training scenario, and all kinds of bodily movements including torso-, and head-posture, muscle tone, blush, breathing, gaze, and eyes openness. For the focus of this article the bodily movements are of particular interest. For the sake of clarity we focus on torso-movement only. We assume postures are described using pitch, role and yaw. However, virtual reflexes can be implemented using another actuator representation.

We focus on the decision node for torso-pitch. Torso-pitch is influenced by the bodily movements of the trainee, but also by the emotion and cognition components. Furthermore, we show that in an indirect way, the Virtual Reflex-loops influence emotion and cognition. We present the formalization of one of the Virtual Reflex loops for the aggression de-escalation case. In particular, we present the contents of the decision node that controls torso pitch. Torso-pitch is the forward or backward angle of the body. It is a key contributor to the approach/avoidance behavior of the virtual character.

An important modeling concept for torso-pitch is the social distance as modulated by emotion (feeling of Dominance of the VC, to be precise). In this example we describe how the torso pitch changes throughout a scenario. We define the dynamics of torso-pitch behavior as follows:

$torso\_pitch_{t+1} = C_{sd}*(sd_{target}-d_t)$  /* normalized for virtual character control */
$sd_{target}=(1-D)*sd_{default}+CD$
$C_{sd}=(A+1)/2$

Here, $C_{sd}$ is the social distance inertia factor [0,1], and $d_t$ is the physical distance at time t. D is Dominance [-1,1] and relates to social stance, perceived control, and self-efficacy. Dominance moderates the default target distance. High dominant Virtual Characters will have a closer preferred social distance, and vice versa. A is Arousal [-1,1] and relates to energy. Energy thus defines the speed with which the torso angle changes. CD is defined as cultural distance. The closer two individuals are with respect to their cultural background, the smaller CD is. CD should be calibrated based on existing social science findings. $Sd_{default}$ is defined as the Virtual Character's personal preferred social distance. This factor is needed to model individual differences in preferred social distance as this varies from person to person. The parameter is set in the cognition model. Similarly, the PAD emotion values are set by the emotion model, thus modeling the effect of emotion on the Virtual Reflex-loop.

Conversely, each of the Virtual Reflex-loops potentially influences emotion and cognition (dotted lines from decision nodes to the emotion component, figure 1). In the case study, the values computed in the torso-pitch decision node are passed to the emotion component. For example, forward torso pitch can induce anxiety or aggression (depending on the VC's personality and the current affective state). Similarly, the cognitive component receives input from the sensors, the decision nodes and the emotion component. For now we focus on the input from the decision nodes that send information on the body state of the VC. The VC's cognition component plans on overall and longer term changes in behavior, e.g., changes in posture, dialogue, etc. The VC's body integrity model ensures that movements caused by different Virtual Reflex-loops result in physically coherent behavior.

## 5.2 Virtual Reflex-based scenario formalization

In this section we describe how torso pitch changes throughout an example scenario in a social welfare office.

1. Behind the social welfare office counter is Barney. Barney is new in his job and feels uncertain. (t=0)
2. A person (the virtual character VC) is frustrated and angry as his allowance has been canceled. He enters the offices with the intention to demand his money. (t=1)
3. VC approaches the counter. (t=2)
4. Barney backs away. (t=3)
5. VC arrives at counter. (t=4)
6. VC leans over the counter (and slams his hand on the counter and shouts "I demand my money!"). (t=5)
7. Barney keeps distance and stays calm eventually calming the VC. (t=6)
8. Social distance is in equilibrium (t=7) VC and Barney have stable distance.

We now describe the effects of the events in the scenario on torso-pitch, as an example of how the Virtual Character behavior is controlled. We assume the VC's body integrity model keeps balance and walks back and forth when the torso is pitched. Therefore, torso-pitch can indirectly control forward/backward walking.

| | |
|---|---|
| **1. Virtual Character (t=0):**<br>  Torso: (0, 0, 0)<br>  Current distance to Barney: undefined<br>  $PAD_{default}$=(0, 0, -0.5) (this represents the VC's personality)<br>  PAD = (-1, 1, 1), $sd_{default}$=1, CD=0.2 (VC is close in culture to Barney) | **4. Barney backs away due to approach VC (t=3)**<br>  Torso: (-1.8, 0, 0)<br>  Current distance to Barney 2.5 meter (larger because Barney backs away)<br>  *Effect on Torso_pitch*:<br>  torso_pitch =$C_{sd}$*($sd_{target}$-$d_t$)=1*(0.7-2.5)= -2.3<br>  $sd_{target}$=(1-D)*$sd_{default}$+CD=(1-1)*1+0.2=0.2<br>  $C_{sd}$=(A+1)/2=1<br>  *Torso is tilted further forward due to Barney moving back, increasing approaching speed of VC.* |
| **2. The VC is positioned inside the room (t=1).**<br>  Torso: (0, 0, 0)<br>  Current distance to Barney: 4 meter<br>  *Effect on torso_pitch*:<br>  torso_pitch =$C_{sd}$*($sd_{target}$-$d_t$)=1*(0.2-4)= -3.8<br>  $sd_{target}$=(1-D)*$sd_{default}$+CD=(1-1)*1+0.2=0.2<br>  $C_{sd}$=(A+1)/2=1<br>  *Torso is tilted forward heavily, resulting in imbalance and a rapid move forward.* | **5. VC arrives at the counter (t=4)**<br>  Torso: (-2.3, 0, 0)<br>  Current distance to Barney 1.5 meter (counter width=1m, Barney backed away 0.5m)<br>  *Effect on Torso_pitch*:<br>  torso_pitch =$C_{sd}$*($sd_{target}$-$d_t$)=1*(0.2-1.5)= -1.3<br>  $sd_{target}$=(1-D)*$sd_{default}$+CD=(1-1)*1+0.2=0.2<br>  $C_{sd}$=(A+1)/2=1<br>  *Torso is tilted forward a little, and because VC arrived at counter, legs can't move, so torso starts to lean over the counter.* |

| 3. The VC approaches the counter (t=2) | 6. VC leans over the counter (t=5) |
|---|---|
| Torso: (-3.8, 0, 0) <br> Current distance to Barney 2 meter (and decreasing quickly) <br><br> *Effect on Torso_pitch*: <br> torso_pitch =$C_{sd}$*($sd_{target}$-$d_t$)=1*(0.2-2)= -1.8 <br> $sd_{target}$=(1-D)*$sd_{default}$+CD=(1-1)*1+0.2=0.2 <br> $C_{sd}$=(A+1)/2=1 <br> *Torso is tilted forward, still resulting in imbalance and a move forward, but slower than before, still VC shows no clear intention to stop moving.* | Torso: (-1.3, 0, 0) <br> Current distance to Barney 1.5 <br><br> *Effect on Torso_pitch*: <br> torso_pitch =$C_{sd}$*($sd_{target}$-$d_t$)=1*(0.2-1.5)= -1.3 <br> $sd_{target}$=(1-D)*$sd_{default}$+CD=(1-1)*1+0.2=0.2 <br> $C_{sd}$=(A+1)/2=1 <br> *VC leans over the counter towards Barney.* |

This short example shows how behavior can be generated by reflex nodes, and how this behavior can be modulated with emotion. It does not show how emotion emerges from the reflex nodes. However, the system of equations should be interpreted as a dynamic system. The VC's Dominance results from interaction with the virtual character as well, simply by the fact that the system settles only at a particular close social distance if dominance is high. (Note the bidirectional nature of the reflex nodes and the emotional state.). So, if the trainee (Barney) is able to calm down the VC (e.g., by keeping distance and staying calm so that the PAD state will decay to $PAD_{default}$, i.e., the personality of the VC), dominance of the VC has to drop, resulting in a larger desired social distance of the VC $sd_{target}$. Arousal also drops due to calm sensory input that modulates $C_{sd}$. This would result in:

| 7. VC calms down (t=6) | 8. Social distance in equilibrium at $sd_{target}$ (t=7) |
|---|---|
| Torso: (-1.2, 0, 0) <br> Current distance to Barney 1.5 <br> PAD=(0,0,-0.5) (state decayed to $PAD_{default}$ due to keeping distance and staying calm) <br><br> *Effect on Torso_pitch*: <br> torso_pitch =$C_{sd}$*($sd_{target}$-$d_t$)=0.5*(1.7-1.5)= 0.1 <br> $sd_{target}$=(1-D)*$sd_{default}$+CD=(1- -0.5)*1+0.2=1.7 <br> $C_{sd}$=(A+1)/2=0.5 <br> *VC leans backwards slightly and thus starts to move back from the counter a bit.* | Torso: (-1.2, 0, 0) <br> Current distance to Barney 1.7 <br><br><br><br> *Effect on Torso_pitch*: <br> torso_pitch =$C_{sd}$*($sd_{target}$-$d_t$)=0.5*(1.7-1.7)= 0 <br> $sd_{target}$=(1-D)*$sd_{default}$+CD=(1- -0.5)*1+0.2=1.7 <br> $C_{sd}$=(A+1)/2=0.5 <br> *VC and Barney have stable distance. Should Barney choose to move forward, the VC will react by moving backward, so they are "in synch".* |

We have not shown how cognition modulates these processes, but any higher-level processing can reason upon and influence these reflexes. The effectiveness for generating complex behaviors in line with a scenario is part of current work.

## 6 Conclusions

The purpose of our work is to achieve immersive and realistic virtual environments for social skill training. To realize this, we propose a new computational model for virtual character behavior. Instead of following the cumbersome route of deducing high-level conceptual emotions from low-level observations, processing them, computing responses at a conceptual level, and translating these into muscle actuations, we propose a neurologically-inspired direct approach. Our low level approach is based on the idea of virtual reflexes, in which observations directly cause muscle actuations *without intermediate cognitive processing*. Emotions emerge out of this interaction between sensory and motor information. Although no cognitive intermediate processing takes place, cognition does modulate the sensory-motor loop. As happens in the human body, various virtual reflexes can occur simultaneously. Therefore, in the paper we have modeled a set of virtual reflexes as concurrent subsystems. This paper is innovative in three ways:
- We introduce an architecture for virtual reflexes.
- We link the architecture to neuropsychological theories on emotion & cognition.
- We formalize part of the reflexes in a virtual aggression training case study.

A side effect of our approach is that the application of this design in a VRET, such as for de-escalating verbal aggression, will be a feasibility test of Damasio's theory of emotion regulation (Damasio, 1999; Damasio & Carvalho, 2013): reaction comes before feeling, and, "feeling the feeling" is emotion.

## Acknowledgements

We gratefully acknowledge the input of Guntur Sandino (CLeVR), Arnaud Wirschell (ROI), Fred Schrijber (aggression trainer), and Ron Knaap (Trigion). This work has further benefited from discussions with Otto Adang, Ron Boelsma, Willem-Paul Brinkman, Koen Hindriks, and Birna van Riemsdijk. We gratefully acknowledge the support of Jaap van den Herik.

## References

1. Argyle, M., (2009). Social Interaction, Transaction Publishers Rutgers, New Jersey.
2. Berthoz, A., (2002). The brains sense of movement. Harvard Univ. Press.
3. Broekens, J., Harbers, M., Brinkman, W.-P., Jonker, C., Bosch, K., & Meyer, J.-J. (2012). Virtual Reality Negotiation Training Increases Negotiation Knowledge and Skill. In Y. Nakano, M. Neff, A. Paiva & M. Walker (Eds.), Intelligent Virtual Agents (Vol. 7502, pp. 218-230): Springer.
4. Brooks, R. A. (1999). Cambrian intelligence: the early history of the new AI: MIT Press.
5. Cafaro, A., Vilhjálmsson, H., Bickmore, T., Heylen, D., Jóhannsdóttir, K., & Valgarðsson, G. (2012). First Impressions: Users' Judgments of Virtual Agents' Personality and Interpersonal Attitude in First Encounters. In Y. Nakano, M. Neff, A. Paiva & M. Walker (Eds.), Intelligent Virtual Agents (Vol. 7502, pp. 67-80): Springer Berlin Heidelberg.
6. Cañamero, L. 2005. Designing Emotional Artifacts for Social Interaction: Challenges and Perspectives. In L. Cañamero, R. Aylett (Eds.), *Animating Expressive Characters for Social Interaction*. Adv in Consciousness Research. Cavazza, M., Charles, F., & Mead, S. J. (2002.). Character-Based Interactive Storytelling. IEEE Intelligent Systems, 17(4), 17-24.
7. Core, M., Traum, D., Lane, H. C., Swartout, W., Gratch, J., van Lent, M. (2006). Teaching Negotiation Skills through Practice and Reflection with Virtual Humans. SIMULATION, 82(11), 685-701.

8. Damasio, A., (1999). The Feeling of What Happens: Body and Emotion in the Making of Consciousness, Harcourt.
9. Damasio, A., Carvalho G.B. (2013). The nature of feelings: evolutionary and neurobiological origins" Nature Reviews Neuroscience 14, 143-152 (February, 2013) PubMed
10. D'Mello, S., Picard, R. W., & Graesser, A. (2007). Toward an Affect-Sensitive AutoTutor. 22, 53-61.
11. Emmelkamp, P. M. G., Bruynzeel, M., Drost, L., & van der Mast, C. A. P. G. (2001). Virtual reality treatment in acrophobia: a comparison with exposure in vivo. Cyber Psychology & Behavior, 4(3), 335-339.
12. Faloutsos, P., Panne, M. v. d., & Terzopoulos, D. (2001). Composable controllers for physics-based character animation. Paper presented at the Proceedings of the 28th annual conference on Computer graphics and interactive techniques.
13. Finkelstein, S., Ogan, A., Walker, E., Muller, R., & Cassell, J. (2012) Rudeness and rapport: Insults and learning gains in peer tutoring. in Intelligent Tutoring Systems, Springer.
14. Folkman, S., & Lazarus, R. S. (1990). Coping and emotion. In N. L. Stein, B. Leventhal & T. Trabasso (Eds.), Psych and Bio Appr Emo (313–332). Hillsdale, NJ: Erlbaum.
15. Gratch, J., Wang, N., Gerten, J., Fast, E., & Duffy, R. (2007). Creating Rapport with Virtual Agents. In C. Pelachaud, J.-C. Martin, E. André, G. Chollet, K. Karpouzis & D. Pelé (Eds.), Intelligent Virtual Agents (Vol. 4722, pp. 125-138): Springer Berlin Heidelberg.
16. Hays, M. J., Ogan, A., & Lane, H. C. (2010). The Evolution of Assessment: Learning about Culture from a Serious Game. In C. Lynch, K. D. Ashley, T. Mitrovic, V. Dimitrova, N. Pinkwart & V. Aleven (Eds.), IllDef2010, pp. 37-44.
17. Heylen, D., Nijholt, A., & Akker, R. o. d. (2005). Affect in tutoring dialogues. Applied Artificial Intelligence: An International Journal, 19(3), 287 - 311.
18. Huang, L., Morency, L.-P., & Gratch, J. (2011). Virtual Rapport 2.0. In H. Vilhjálmsson, S. Kopp, S. Marsella & K. Thórisson (Eds.), Intell. Virt. Agents 6895:(68-79): Springer.
19. D. Kahnemann. Thinking, fast and slow. Penguin Books, 2011.
20. Kavanagh, L., Bakhtiari, G., Suhler, C., Churchland, P.S., Holland, R.W. & Winkielman, P. (2013). Nuanced Social Inferences about Trustworthiness from Observation of Mimicry. In M. Knauff, M. Pauen, N. Sebanz, & I. Wachsmuth (Eds.),*Proceedings of the 35th Annual Conference of the Cognitive Science Society*. Berlin, Germany: Cognitive Science Society. 734-739.
21. Kendon, A. 1990. Conducting interaction. Studies in Interactional Sociolinguistics, vol. 7. Cambridge University Press.
22. Kim, J. M., Hill, J. R. W., Durlach, P. J., Lane, H. C., Forbell, E., Core, M., et al. (2009). BiLAT: A Game-Based Environment for Practicing Negotiation in a Cultural Context. International Journal of Artificial Intelligence in Education, 19(3), 289-308.
23. Krijn, M., Emmelkamp, P. M. G., Olafsson, R. P., & Biemond, R. (2004). Virtual reality exposure therapy of anxiety disorders: A review. Clinical Psych Review, 24(3), 259-281.
24. LeDoux, J. (1996). The Emotional Brain. New York: Simon and Shuster.
25. LeDoux, J. E. (1995). Emotion: Clues from the brain. Ann. Rev. Psy., 46(1), 209-235.
26. Magnenat-Thalmann, N., & Thalmann, D., (2005). Virtual humans: thirty years of research, what next? The Visual Computer 21: 997-1015.
27. Marsella, S., Gratch, J., Ning, W., & Stankovic, B. (2009, 10-12 Sept. 2009). Assessing the validity of a computational model of emotional coping. Affective Computing and Intelligent Interaction and Workshops, 2009. ACII 2009.
28. Meichenbaum, D. (1994). A clinical handbook/practical therapist manual for assessing and treating adults with post-traumatic stress disorder (PTSD): Waterloo, Canada: Inst. Press.
29. Mehrabian, A. (1980). Basic Dimensions for a General Psychological Theory: OG&H

30. Ortony, A., Clore, G. L., & Collins, A. (1988). The Cognitive Structure of Emotions: Cambridge University Press.
31. Pantic, M., & Rothkrantz, L. J. M. (2003). Toward an affect-sensitive multimodal human-computer interaction. Proceedings of the IEEE, 91(9), 1370-1390.
32. Parsons, S., & Mitchell, P. (2002). The potential of virtual reality in social skills training for people with autistic spectrum disorders. J. Intell. Disability Research, 46(5), 430-443.
33. Popescu, A., Broekens, J., & Someren, M. v. (2013). GAMYGDALA: An Emotion Engine for Games. IEEE Transactions on Affective Computing, 99(PrePrints), 1-1.
34. Popović S1, Horvat M, Kukolja D, Dropuljić B, Cosić K. (2009). Stress inoculation training supported by physiology-driven adaptive virtual reality stimulation, Stud Health Technol Inform. 2009;144:50-4.
35. Powers, M. B., & Emmelkamp, P. M. G. (2008). Virtual reality exposure therapy for anxiety disorders: A meta-analysis. Journal of Anxiety Disorders, 22(3), 561-569.
36. Scherer, K. R. (2001). Appraisal considered as a process of multilevel sequential checking. In K. R. Scherer, A. Schorr & T. Johnstone (Eds.), Appraisal processes in emotion: Theory, methods, research (pp. 92-120).
37. Schroder, M., Bevacqua, E., Cowie, R., Eyben, F., Gunes, H., Heylen, D., et al. (2012). Building Autonomous Sensitive Artificial Listeners. IEEE Aff. Computing, 3(2), 165-183.
38. Serino, S., Triberti, S., Villani, D., Cipresso, P., Gaggioli, A., & Riva, G. (2013). Toward a validation of cyber-interventions for stress disorders based on stress inoculation training: a systematic review. Virtual Reality, 1-15.
39. Sevin, E., Hyniewska, S., & Pelachaud, C. (2010). Influence of Personality Traits on Backchannel Selection. In J. Allbeck, N. Badler, T. Bickmore, C. Pelachaud & A. Safonova (Eds.), Intelligent Virtual Agents (Vol. 6356, pp. 187-193): Springer Berlin Heidelberg.
40. Spek, E. v. d. (2011). Experiments in Serious Game Design. University of Utrecht.
41. Tekofsky, S., Spronck, P., Plaat, A., Van den Herik, J., & Broersen, J. (2013). Psyops: Personality assessment through gaming behavior. Paper presented at the Proceedings of the International Conference on the Foundations of Digital Games.
42. Theune, M., Faas, S., Heylen, D. K. J., & Nijholt, A. (2003). The virtual storyteller: Story creation by intelligent agents TIDSE 2003: Technologies for Interactive Digital Storytelling and Entertainment (pp. 204-215). Darmstadt: Fraunhofer IRB Verlag.
43. Thiebaux, M., Marsella, S., Marshall, A. N., & Kallmann, M. (2008). Smartbody: Behavior realization for embodied conversational agents Proceedings of the 7th Conf, Autonomous agents and multiagent systems. Vol 1 (pp. 151-158): IFAAMS.
44. Wakefield, Jerome C. & Dreyfus, Hubert L. (1991). Intentionality and the phenomenology of action. In Lepore & Gulick (eds.), John Searle and His Critics. Cambridge: Blackwell.
45. Wilson, M. (2002). Six views of embodied cognition. Psych. Bull. & Rev., 9(4), 625-636.
46. Zeng, Z., Pantic, M., Roisman, G. I., & Huang, T. S. (2009). A Survey of Affect Recognition Methods: Audio, Visual, and Spontaneous Expressions. Pattern Analysis and Machine Intelligence, IEEE Transactions on, 31(1), 39-58.
47. Ziemke, T. (2003). What's that thing called embodiment Proceedings of the 25th annual conference of the cognitive science society (pp. 1305-1310): Mahwah, NJ: Erlbaum.
48. Zwaan, J., Dignum, V., & Jonker, C. (2012). A Conversation Model Enabling Intelligent Agents to Give Emotional Support. In W. Ding, H. Jiang, M. Ali & M. Li (Eds.), Modern Advances in Intelligent Systems and Tools (Vol. 431, pp. 47-52): Springer.